# DESIGNING FOR INTERPRETATION UNCERTAINTY: ARCHITECTURE AND PRINCIPLES FOR TOPOLOGICAL LEARNING ANALYTICS DASHBOARDS

Hitoshi Inoue *
*Nakamura Gakuen University*
*5-7-1 Befu, Jonan-ku, Fukuoka City, 814-0198 JAPAN **

Koichi Yasutake **
*Hiroshima University*
*1-1-1 Kagamiyama, Higashi-Hiroshima City, 734-8524 JAPAN ***

**ABSTRACT**

Topological Data Analysis (TDA) offers novel methods for understanding temporal dynamics in complex systems, yet its application in information systems design faces a fundamental challenge: how should systems present analytical outputs when interpretation frameworks are still developing? This paper reports on the development of **TopoLA,** a dashboard system applying Zigzag Persistent Homology to learning management system data, and proposes three early design principles for interpretation support in emerging analytics: (1) separation of objective measurement from contextual interpretation, (2) graduated disclosure from metrics through patterns to reflective prompts, and (3) explicit acknowledgment of methodological uncertainty. The system implements a modular three-stage pipeline—feature extraction, topological computation, and interpretation support—enabling extension to additional analytical methods. This work contributes to information systems research by articulating preliminary design knowledge for systems that must communicate analytical insights from methods lacking established interpretation norms—a challenge increasingly common as novel computational techniques enter applied domains.

## 1. INTRODUCTION

Organizations increasingly adopt novel computational methods to extract insights from complex data. Topological Data Analysis (TDA), machine learning interpretability techniques, and network-based temporal analytics reveal patterns invisible to traditional methods. However, a fundamental design challenge accompanies this innovation: how should information systems present outputs from methods where interpretation frameworks remain underdeveloped?

This challenge differs from conventional dashboard design. When displaying familiar metrics—click rates or assessment scores—designers can assume shared understanding of what values mean. Emerging analytical methods lack this foundation. For example, a topological indicator showing "$\beta_0$=184"—where $\beta_0$ represents the number of connected components, or behaviorally similar learner groups—conveys precise mathematical meaning (184 distinct behavioral clusters), yet its practical significance—whether this represents problematic fragmentation or healthy diversity—remains context-dependent and theoretically contested.

Recent systematic reviews have documented this challenge across the learning analytics field. Kaliisa et al. (2024) found that dashboards predominantly increase awareness "but with limited actionable insights," while Schwendimann et al. (2017) noted that most dashboard research focuses on conventional visualizations—bar charts, line graphs, and tables—derived from familiar metrics. Jivet et al. (2017) specifically identified that "awareness is not enough" and that educators struggle to translate dashboard information into pedagogical action. These findings suggest a systematic gap between analytical capability and interpretive support.

Published in the Proceedings of the IADIS International Conference Information Systems 2026, pp. 506-510. Copyright © IADIS. Posted with permission under the rights retained by the authors in Clause 4 of the IADIS Transfer of Copyright Agreement, which permits distribution as a preprint, reprint, or related class of document.

The challenge intensifies when analytical methods lack established interpretation frameworks. While uncertainty visualization research addresses quantified uncertainty (Padilla et al., 2022), such approaches assume the interpretation framework itself is settled. We distinguish this as "interpretive uncertainty": uncertainty not about precision of values, but about the framework for understanding what those values mean in practice.

This paper reports on the development of TopoLA, a dashboard implementing Zigzag Persistent Homology for learning management system data. Through iterative prototyping following Design Science Research methodology (Hevner et al., 2004; Peffers et al., 2007), we derived three early design principles addressing the tension between providing actionable guidance and respecting methodological limitations. Our contributions are: (i) a modular three-stage architecture enabling method extensibility, (ii) a three-layer interpretation support interface implementing the proposed principles, and (iii) identification of evaluation directions for future validation. While demonstrated in learning analytics, these principles may generalize to other contexts where novel computational techniques enter applied practice.

## 2. BACKGROUND AND RELATED WORK

Topological Data Analysis (TDA) applies algebraic topology to extract structural features from high-dimensional data (Carlsson, 2009; Chazal and Michel, 2021). Zigzag Persistent Homology extends standard persistence to non-monotonic sequences, enabling temporal analysis (Carlsson and de Silva, 2010). The key indicator $\beta_0$ represents connected components at a given threshold—capturing how learners cluster into behaviorally similar groups. Translating these constructs into domain-specific interpretations remains challenging (Munch, 2017).

Systematic reviews reveal persistent challenges in dashboard design. Kaliisa et al. (2024) concluded from meta-analysis that learning analytics dashboards have not consistently improved outcomes. A key issue is inadequate interpretation support—most dashboards increase awareness "but with limited actionable insights" (Kaliisa et al., 2023). Visual representations remain "limited to graphs, charts or other diagrams without providing support mechanisms to facilitate their interpretation" (Sedrakyan et al., 2018). Ahn et al. (2019) emphasize that dashboard design must be deeply embedded in the contexts of use, attending to practitioners' data sensemaking needs beyond mere usability. Verbert et al. (2013) and Schwendimann et al. (2017) document that while dashboard applications proliferate, systematic approaches to interpretation support remain underdeveloped.

While uncertainty visualization research addresses confidence communication in mature domains (e.g., Padilla et al., 2022), such work focuses on quantified uncertainty—probability distributions, confidence intervals, error bounds. Emerging analytics present a distinct challenge: interpretive uncertainty, where the framework for understanding what outputs mean remains contested, not merely the precision of values. Current dashboard approaches exhibit two problematic patterns: presenting metrics without guidance (assuming user self-sufficiency) or embedding definitive interpretations (overstating analytical certainty). With emerging analytical methods lacking established interpretive foundations, both failure modes become more severe—users cannot self-interpret unfamiliar metrics, yet definitive guidance lacks empirical grounding.

## 3. SYSTEM ARCHITECTURE

TopoLA implements a modular three-stage pipeline. The first stage transforms LMS logs into behavioral feature vectors. Drawing on Activity Theory (Engeström, 1987), the implementation uses ten dimensions capturing material-type diversity, VLE interactions, assessment metrics, and submission patterns—ensuring features capture educationally meaningful constructs rather than arbitrary statistical dimensions. All features undergo global z-standardization.

The second stage applies Zigzag Persistent Homology to weekly behavioral snapshots using *Dionysus* 2.x (Morozov, 2008). Proximity graphs connect learners whose Euclidean distance falls below threshold $\varepsilon = 1.5$, selected through calibration targeting 10–20% isolated learners. Sensitivity analysis across $\varepsilon \in [1.0, 2.0]$ confirmed $\beta_0$ patterns remain stable (Spearman $\rho > 0.85$).

The third stage, built with *Streamlit*, a Python-based web application framework, presents outputs through the interpretation support framework described below.

Central to this presentation are two composite indicators derived from $\beta_0$ dynamics, designed for educator accessibility:

The Topological Early Warning Index ($TEWI$) quantifies structural fragmentation as the rate of change in $\beta_0$ over a sliding window:

$$TEWI_t = (\beta_0(t) - \mu_w(t)) \,/\, \mu_w(t)$$

where $\mu_w(t)$ represents the mean $\beta_0$ over the $w$ weeks preceding week $t$ (default $w$ = 3). Higher $TEWI$ values indicate rapid fragmentation increase warranting attention. This formulation captures sudden structural changes relative to recent baseline, enabling timely detection of emerging fragmentation patterns.

The Trajectory Stability Index ($TSI$) measures individual learner consistency:

$$TSI_i = 1 - CV(x_i)$$

where $CV(x_i)$ represents the coefficient of variation—the ratio of standard deviation to mean—of learner $i$'s weekly behavioral vectors across time. Higher $TSI$ indicates more stable, consistent learning patterns.

Both indicators are normalized 0–1 within each course across all weeks, enabling temporal comparison within courses while acknowledging cross-course comparison limitations.

## 4. EARLY DESIGN PRINCIPLES FOR INTERPRETATION SUPPORT

Through iterative prototyping, we identified three preliminary design principles addressing communication of analytical outputs from methods with underdeveloped interpretation frameworks. TopoLA is designed as an exploratory sensemaking tool rather than a prescriptive diagnostic system. Designers face a fundamental tension: providing no interpretive guidance leaves users unable to translate outputs into action, while providing definitive interpretations overstates methodological maturity and may lead to inappropriate actions. Effective design must bridge this asymmetry—leveraging user contextual expertise while acknowledging methodological limitations.

**Principle 1:** Separation of Measurement from Interpretation. Present objective analytical outputs independently from interpretive guidance, enabling users to distinguish empirical findings from suggested meanings. Anti-pattern to avoid: labeling metrics directly as "Healthy" or "At-risk" without separation. In TopoLA, quantitative metrics (e.g., "$TEWI$: 0.318", "$\beta_0$: 184") appear in dedicated interface regions with provisional attention cues (colors signal attention level, not definitive status), visually distinct from interpretive content, so users encounter raw outputs before any suggested interpretation.

**Principle 2:** Graduated Disclosure. Structure interpretive support as progressive layers, moving from objective metrics through research-grounded descriptions to reflective prompts inviting contextual interpretation. Anti-pattern to avoid: presenting only recommended actions without underlying evidence. TopoLA implements three expandable layers: Layer 1 presents numerical indicators without interpretation; Layer 2 offers tentative observations using hedged language (e.g., "Values below 0.4 have been associated with relatively cohesive patterns in pilot analyses; thresholds may vary by context"); Layer 3 provides suggested actions inviting contextual application (e.g., "Consider promoting discussion in forums").

**Principle 3:** Explicit Uncertainty Acknowledgment. Communicate methodological maturity transparently through language choices, visual design, and explicit statements about interpretive limitations. Anti-pattern to avoid: relegating uncertainty to footnotes while presenting confident conclusions in the main interface. TopoLA employs multiple mechanisms: hedged language throughout, exploratory framing, and role-based view adaptation—Teachers see Layers 1–2 expanded by default with Layer 3 collapsed but accessible; Designers access trend analyses; Researchers see all layers expanded plus parameter controls and raw data exports.

These principles address varying levels of analytical literacy among end users. The role-based view adaptation ensures users encounter information calibrated to their expertise, aligning with cognitive load theory (Sweller, 1988) and findings that dashboards often overwhelm users with unfiltered data (Jivet et al., 2017). By separating metrics from interpretation, users with different backgrounds can engage productively.

These principles operate synergistically: separation establishes the foundation; graduated disclosure structures support; uncertainty acknowledgment maintains epistemic humility. Any domain adopting novel computational techniques—explainable AI, network-based social analytics, emerging bioinformatics—faces analogous challenges. These principles offer preliminary design knowledge for responsible integration across such contexts.

## 5. IMPLEMENTATION AND PLANNED EVALUATION

We demonstrate the architecture using the Open University Learning Analytics Dataset (OULAD), containing behavioral traces from 22 courses with approximately 32,000 learners (Kuzilek et al., 2017). The dashboard is implemented in Python, with research analysis code from Jupyter Notebooks directly integrated into the web application—ensuring consistency between research findings and dashboard outputs.

Future evaluation will explore how educators interact with the three-layer interface, focusing on whether the graduated disclosure supports appropriate interpretation without overwhelming users. A key outcome measure will be overconfidence rate—the proportion of high-confidence interpretations paired with incorrect judgments.

Figure 1 illustrates the three-layer interpretation support described in Section 4.

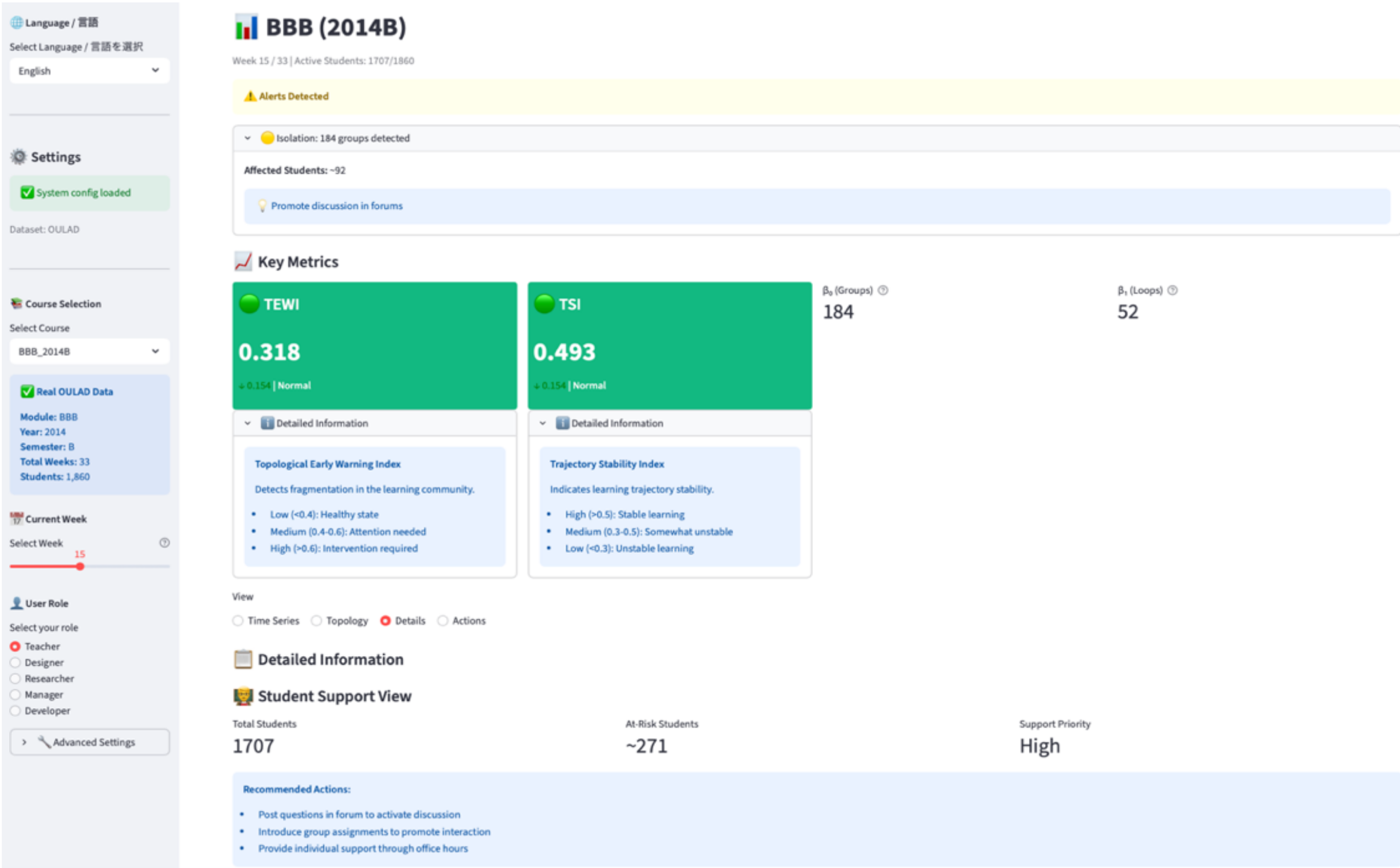

Figure 1. TopoLA dashboard interface demonstrating three-layer interpretation support.

## 6. DISCUSSION AND CONCLUSION

This paper contributes preliminary design knowledge at two levels. First, TopoLA demonstrates TDA integration into dashboard systems with a modular architecture. Second, we propose three design principles—separation of measurement from interpretation, graduated disclosure, and explicit uncertainty acknowledgment—for systems communicating insights from methods lacking established norms. These principles address challenges shared across domains where novel computational techniques outpace interpretive consensus, offering initial guidance for responsible system development.

Limitations and Future Work. Several important limitations constrain current claims. First, the design principles emerged from single-domain development (learning analytics) without formal user evaluation; their generalizability to other domains adopting novel computational techniques remains theoretical. Second, the TEWI and TSI indicators, while grounded in topological theory, lack empirical validation linking specific threshold values to learning outcomes—we demonstrate structural patterns but not predictive utility. Third, the current implementation uses OULAD data from a specific institutional context; cross-institutional validation is essential before practical deployment.

Future directions include empirical validation of the proposed evaluation framework, integration with established learning analytics methods (transition network analysis (Saqr et al., 2026), sequence mining), and application to other domains where novel computational techniques outpace interpretive consensus—particularly explainable AI systems and network-based social analytics.

## ACKNOWLEDGEMENT

This work was supported by JSPS KAKENHI Grant Numbers JP25K00845, JP25K21951, JP23K22315, and JP23K17619.